\documentstyle{article}[12pt]
\date{}

\begin{document}
\title{Non-Locality and Theories of Causation
\thanks{Presented at the NATO Advanced Research Workshop \emph{Modality, Probability and Bell's
Theorems}, Cracow, Poland, August 19-23, 2001.}
}
\author{Federico Laudisa\\
Faculty of Sciences of Education, University of Milan-Bicocca\\
Piazza dell'Ateneo Nuovo, 1, 20126 Milan Italy\\
\tt laudisa@mailserver.idg.fi.cnr.it}
\maketitle
\date{}
\noindent
\\

\section{Introduction}

In the natural as well as in the social or psychological domain, puzzling phenomena call for
an explanation, and there is little doubt that the connection among quantum events across
spacetime - known as non-locality - is indeed puzzling. Events that we might reasonably
consider mutually independent, according to our best theory of space and time, turn out
to influence each other. But as soon as we try to understand what this `influence' could
amount to, we find ourselves in deep physical and philosophical troubles, and if we attempt
to investigate the connection between non-locality and causation, the situation may become
even more complicated. For if for the sake of the argument we assume we have a vague intuition
of what non-locality might be, several are the questions worth asking. Is a causal view of
non-locality itself possible? In particular, can the nature of quantum non-locality be
somehow clarified by viewing it as grounded in some (perhaps unfamiliar) sort of causation?
Which properties should this sort of causation satisfy?

There are two preliminary and general circumstances that need to be taken into account but that,
at the same time, contribute to make the picture unclear. First, there seem to be different
ways in which non-locality is manifested in quantum mechanics. Second, the notion of causation
itself is far from being understood in an univocal and uncontroversial sense. The intuition
according to which the occurrence of a physical event $A$ determines (produces, brings about,
raises the probability of, ...) the occurrence of a distinct physical event $B$
- in which case $A$ is said to be the `cause' of  $B$ - can be represented differently in
different causal theories. Within the physicists' community, for instance, it is assumed -
tacitly or not - that events recognized to be causes must be temporally prior to their alleged
effects, and the causal doctrine based on this assumption is sometimes referred to as
`relativistic causality'. This terminology is itself biased, however, since it takes for
granted that special relativity provides the strongest possible support for this assumption.
In fact, a rich philosophical debate has shown that if, more generally, the only requirement
to be satisfied is the impossibility of generating causal paradoxes, several causal theories
may be developed without assuming any temporal priority of causes. Moreover, different causal
theories may have a differing degree of adequacy when applied to the domain of microphysics.
The evaluations that may be made of their basic causal principles according to different
formulations and applications of the principles themselves may widely differ, so that when
one claims to defend or counteract a causal view of non-locality, he should specify in advance
what is the causal theory in terms of which that view is supposed to be `causal'. A clear
demonstration of the interpretation-dependent character of causal notions is the debate on
Reichenbach's common cause principle, according to which when two events A and B are
correlated, either there is a direct connection between A and B producing the correlation or
there is a different event C which causes the correlation. On the basis of different
intuitions and formal definitions, opposite conclusions have been drawn on whether
explanations of nonlocal quantum correlations in terms of probabilistic common causes are
an option or not. This circumstance strongly supports in my opinion the view expressed in [8],
according to which what is usually called the common cause principle ``is not really a principle
but a schema of principles that calls for interpretation'' (p. 53).
\footnote{For a recent and general assessment of this issue, in addition to [8], cf. the
chapter 3 of [28].}
The pluralism of formulations that both the notion of (non-)locality and the notion of
causation may assume in different theoretical frameworks can be considered primarily as
a \emph{logical} problem. In the assessment of the status and significance of a causal view
of non-locality, however, we have first to take into account its \emph{physical} background, namely
we have to take into account the investigations on the physical meaning of non-locality in
quantum mechanics. The standard framework is that of EPR-Bell correlation experiments,
involving a two spin-1/2 particles' system $S_1+S_2$ prepared in the singlet state, and such
that the spin measurements are performed when the two subsystems $S_1$ and $S_2$ occupy two
space-like separated spacetime regions $R_1$ and $R_2$, respectively, after leaving the source.
The common feature of these investigations is basically an assumption of incompleteness for
the purely quantum description of physical states; on the basis of such assumption a `finer'
state description is postulated via the introduction of extra (`hidden') variables that `add up'
to the quantum state. In this vein the first step was to introduce deterministic hidden
variable models, in which the source state $\lambda$ is postulated to be complete and assumed
to determine with certainty the outcome of any measurement that can be performed on the two
distant subsystems. Later the condition of determinism for hidden variables has been relaxed.
Stochastic hidden variable models were then introduced, in which the state description $\lambda$
allowed by the model enables one to determine not the measurement outcome but only its
probability of occurrence.

Both in the deterministic and stochastic frameworks, a locality condition is usually motivated
by a prescription of `lack of influence' between the spacetime regions in which the
measurement events are localized, although the specific condition of locality that was assumed
in deterministic hidden variables models had to be reformulated in order to comply with the
stochastic character of the more general model. The locality condition was then formulated
as an independence constraint on the statistical predictions generated by the complete
descriptions of the single particles' states (when the particles themselves are spatially well
separated). Namely, the assumption of the mutual independence between the relevant spin
measurement events was formulated as the invariance of the probabilities prescribed by $\lambda$
for any outcome in one wing of the experiment under the change of some relevant parameter in
the distant wing. Consequently, several discussions focused on what different locality
conditions obtained when such parameter was taken to represent different things,
typically parameters pertaining either to apparatus settings or to outcomes of the
measurements. It is worth emphasizing that I refer here to hidden variables \emph{models},
and not to hidden variables \emph{theories}, for a simple reason. In the history of the
hidden variables' issue, the `theories' in which more and more general locality conditions
were assumed - and whose predictions have been shown to be inconsistent with those
of quantum mechanics - were in fact theories only as a \emph{fa\c{c}on de parler}; whereas
the only full-fledged formal construction deserving the title of theory, namely Bohmian
mechanics, is explicitly nonlocal.
\footnote{A recent detailed analysis of these and related issues is in [5,6].}

The greater generality of these stochastic hidden variables models should make the
conclusions drawn from them stronger. If locality is violated in these models, the existence
of non-local influences is strongly supported, and thus their significance for the notion of
causation can be investigated. However, even this more general framework provides no clear
answer to the following central questions:

\noindent
(a) How should the causal meaning of non-locality be assessed by the point of view of
the spacetime structure in which non-local correlations display themselves?

\noindent
(b) Provided we adopt the most natural interpretation of probability in physics, namely the
relative frequency interpretation, and we do not turn to highly controversial notions such
as chances, propensities or dispositions, what might non-local correlations tell us about
single events confined in bounded spacetime regions?
\footnote{In [13] Dickson has questioned the adequacy of locality conditions based on
probabilistic independence when Bohmian mechanics is taken into account, and he argued
that Bohmian mechanics may be shown to satisfy or violate that kind of locality
depending on how a specific model of the theory is constructed This indicates,
according to Dickson, that probabilistic independence is not adequate to capture
the meaning of locality. It is worth recalling that the Dickson argument concerning
the status of locality as probabilistic independence in Bohmian mechanics has been
challenged in [26].}
This is why in the sequel, when I will discuss the status and significance of causal relations
within the issue of non-locality in quantum mechanics, I will assume as a working hypothesis
that causal relations may be analyzed as holding among single events in spacetime, on the
basis of processes that need not refer to any recurrence in order to be considered `causal'.
As every philosopher of causation will immediately acknowledge, this assumption is somehow
reminiscent of a \emph{singularist} approach to causation, endorsed among others by such eminent
philosophers as C.J. Ducasse and G.E.M. Anscombe. In the singularist view of causation
\begin{quotation}
the cause of a particular event [is defined] in terms of but a single occurrence of it,
and thus in no way involves the supposition that it, or one like it, ever has occurred before
or ever will again. The supposition of recurrence is thus wholly irrelevant to the meaning of
cause; that supposition is relevant only to the meaning of law. And recurrence becomes related
at all to causation only when a law is considered which happens to be a generalization of facts
themselves individually causal to begin with. [...] The causal relation is essentially a
relation between concrete individual events; and it is only so far as these events exhibit
likeness to others, and can therefore be grouped with them into kinds, that it is possible
to pass from individual causal facts to causal laws. ([15], pp. 129-30).
\end{quotation}

I wish to stress, however, that I am not embracing a preliminary philosophical position on
causation, namely singularism, and then turning to argue that causation in quantum
mechanics can only make sense if interpreted in singularist terms. As I will discuss more in
detail later, non-locality in quantum mechanics involves a fundamental reference to
counterfactual situations, and since non-trivial counterfactuals are usually supposed
to be grounded in laws supporting them, an orthodox singularist might be already suspicious.
The meaning I attach to singularism is rather general and so is the motivation for adopting
such a viewpoint. If for the sake of my investigation I admit the \emph{a priori} possibility of
discovering a totally new form of causation, that might explain the `action at-a-distance'
allegedly entailed by non-locality (I briefly review the modalities of such `action'
in section 2), I still conceive it to involve physical processes connecting single events.
That is, I incline to interpret this hypothetical causation as a sort of singular phenomenon,
that is enhanced by the actualization of a property instantiated by a physical event and that
affects the actualization of different properties pertaining distant events. The causal action
displayed by this phenomenon should thus be understood as taking place in spacetime in some
well-specified sense, although clearly not as a process propagating continuously in spacetime
([7]). So the question is: how and to what extent can this unfamiliar causation be interpreted
consistently with the more familiar spacetime structure in  which - according to our
well-established physical theories - single physical events live?

Within ordinary quantum mechanics - namely quantum mechanics with state reduction -
a reasonable starting point for addressing the problem is in my opinion is to consider
the implications of this singularist view on non-locality and causation when the state
reduction is taken into due account. In the usual interpretation of quantum mechanics,
state reduction is not only included among the basic postulates of the theory but is
also assumed to be a real physical process. In this interpretation, it is state reduction
that is supposed to actualize most properties of quantum systems, and this is a very general
motivation for pursuing an analysis of the conceptual link between causation and state
reduction. But there is also a more specific motivation for the study of such link.
The events that might be causally connected are assumed to be located at space-like
separated regions: thus if we take seriously - as we should - the spacetime geometry
that underlies this assumption (something that Maudlin calls the \emph{relativistic constraint}:
see [25], pp. 290-2), then we also have to take into account at least some ways out of the
problem of the non-covariance of the state reduction process in relativistic quantum mechanics.
In particular, in view of this problem, section 3 is devoted to the exploration of some of the
implications that different assumptions on where the state reduction occurs may have on the
link causation-reduction.
\footnote{ For the sake of the present discussion, I assume such notions as property
or emergence as uncontroversial. Of course they are not, but in my opinion it is
anyway doubtful that a purely philosophical analysis of such notions could substantially
contribute to a better understanding of the main issues in the foundations of quantum
mechanics.}
In following this line of analysis I do not assume, however, that a causal view of non-locality
cannot be evaluated in a quantum theory without state reduction. Although for obvious reasons
I will not take into account all no-collapse interpretations of quantum mechanics, in
section 4 I will consider what might be the place occupied by causation in standard Bohmian
mechanics. Finally, some tentative conclusions on the prospects of a causal view of
non-locality are summarised.

\section{Non-Locality, Superluminal Dependence and Causation}

Having reasons to believe that, given two events $A$ and $B$, their occurrences depend on
(or influence or affect) one another, is not sufficient in general to claim that $A$ and $B$
are causally connected. On the other hand, a mutual dependence between $A$ and $B$ is a good
reason for us to search whether such dependence is grounded in some underlying causal
mechanism, so far unknown to us. In the context of the EPR-Bell correlations in quantum
mechanics, the events under consideration are assumed to be space-like separated,
so that the search for causation in this context is a search for a superluminal causation,
pursued under the assumption that our quantum-mechanical events display at least a
superluminal dependence.

In order then to investigate whether long distance correlations in EPR-Bell experiments
deserve to be called causal, it is convenient to briefly review the reason why in ordinary
quantum mechanics such correlations can be in fact regarded as an instance of superluminal
dependence between events that in a purely relativistic perspective should be taken to be
mutually independent. For the sake of simplicity, I will assume here that performing a
measurement and detecting an outcome are not distinct events: the terms of the hypothetical
causal connection that I wish to investigate are then to be meant as
\emph{measurement-and-outcome events}.

In a standard EPR-Bell correlation experiment involving a two spin-1/2 particles' system
$S_1+S_2$ prepared in the singlet state, we know that the spin measurements are supposed to
be performed when $S_1$ and $S_2$ occupy two space-like separated spacetime regions $R_1$ and
$R_2$, respectively. Under the hypothesis that quantum predictions are correct, $S_1$ and $S_2$
exhibit a perfect spin correlation, namely if the outcome of an actual measurement of the spin
up along any direction $x$ for the particle $S_1$ is $+1$, the probability of obtaining $-1$
as outcome of the measurement of the spin up along the direction $x$ for the particle $S_2$
equals $1$. Hence, we may say that had the measurement of the spin up along any direction $x$
for the particle $S_1$ come out $-1$, we would have obtained with certainty $+1$ for $S_2$.
However, in ordinary quantum mechanics the measurement process is stochastic, namely from
identical preparations we may obtain different outcomes: the spin of $S_1$ can be either $+1$
or $-1$ in different runs also when the whole set of events causally relevant to obtaining $+1$
or $-1$, localized in the backward light cone of the that event, is exactly the same. But
if $S_1$ and $S_2$ are shown to be perfectly correlated in their outcomes, either there
is a direct dependence between the two measurements, performed in the space-like
separated regions $R_1$ and $R_2$, or there is an dependence between the measurement of
$S_1$ $[S_2]$ and some event in the backward light cone of $S_2$ $[S1]$, and in both cases
the dependence holds between space-like separated events, namely it is superluminal
(cfr. [24], pp. 128-136, and [25], pp. 285-289). Moreover, due to the Bell theorem, any theory
assuming the existence of events or factors that (i) are causally relevant to obtaining $+1$
or $-1$ for $S_1$ $[S_2]$, (ii) are located in the backward light cone of $S_1$ $[S_2]$, and
(iii) screen off the causal relevance which is in the backward light cone of $S_1$ $[S_2]$
(but not in the overlap of the backward light cones of $S_1$ and $S_2$), is bound to give
predictions that disagree with those of quantum mechanics.
\footnote{As should be clear from the above account, the stochastic nature of the measurement
process makes the instance of superluminal dependence even more perspicuous. On the
difficulties of making sense of locality - and of the superluminal dependence that
its violation would imply - in a strictly deterministic theory, see [13], [14] and [26].}

Before going on, a pair of remarks concerning possible objections to the above argument in favor
of superluminal dependence. First, it is worth stressing that in the above argument
counterfactuals are involved just to express the content of the spin strict correlation
property, whereas the locality condition that is presupposed is expressed in terms of the
invariance - across possible different runs of the experiment - of the light cone structure
of the events that are causally relevant to obtaining a given outcome. The latter condition
is independent in principle from any sort of counterfactual locality condition, such as
`the outcome of a measurement on $S_2$ of the spin $x$-component would have been still $+1$,
had the spin component  been measured on $S_1$ in the $z$-direction instead of the $x$-direction',
a formulation which is exposed to the objection of having non-contextuality tacitly built in:
in fact there is no reason why the measurement on $S_2$ of the spin component in a given
direction should have the same outcome when in different runs of the experiment it is measured
with observables of $S_1$ that are mutually incompatible. Second, the above argument does not
rely on the assumption that, after a measurement has been performed and an outcome obtained,
there is necessarily a value of the measured system that corresponds to the outcome
(and hence that, after the completion of the measurement, the measured system satisfies
the definite property of having that value). Namely, the argument holds also if we just assume
that, after the completion of the measurement, the outcomes $+1$ and $-1$
are definite properties \emph{of the measuring apparatuses}.

The relation between outcomes of spin measurements in EPR-Bell correlation experiments is then
an instance of superluminal dependence. Such terms as `dependence' or `influence' are
admittedly vague, however, so that the attempt to elaborate arguments by which we could
legitimately interpret superluminal dependence as a form of causation appear at first
completely reasonable. In addition, there are already well-developed theories of causation
at our disposal, and in principle we are able to analyze the viability of their main
assumptions and conditions by the particular viewpoint of the nature of the dependence
between distant quantum events.

According to Maudlin's terminology, for instance, correlated events like the outcomes of
EPR-Bell correlation experiments are \emph{causally implicated with each other}, a formulation
that is supposed to suggest that the causal implication need not distinguish causes from
effects, and it may hold between events neither of which is a direct cause of the other ([24]).
The generality of the definition has a non trivial justification. If we decide to adopt or
develop a more sophisticated theory of causation, in which more stringent conditions on the
identification of causes and effects are required, we immediately run into difficulties: due to
the space-like separation between the dependent events, the time ordering between them is
non-invariant across different Lorentz frames.
A first option is trying to dissolve the problem, rather than solving it, by arguing that
the very distinction between cause and effect is hardly applicable to EPR-Bell frameworks.
This position, albeit logically consistent, seems to imply that we do not need even
to stipulate what are the terms of the allegedly causal relation that we are investigating.
I will not discuss this option further since I doubt that anything relevant to a decent
notion of causation is left in it. A second option is to retain the distinction between
cause and effect, but to argue that it is the very time ordering associated to \emph{any}
Lorentz frame that \emph{defines} which is the cause and which the effect. In this option
the cause-effect distinction is thus not rejected but is remarkably weakened, since
it acquires itself the status of a frame dependent distinction.

No matter which of the first two preceding options is adopted, however, the superluminal
dependence between EPR-Bell outcomes appears to be `causal' in such a weak sense as to prompt
the question: having acknowledged that EPR-Bell outcomes are somehow connected across
spacetime, do we really obtain any deep insight by calling `causation' that connection?
Or rather what we are doing when we say that the EPR-Bell outcomes are `causally implicated
with each other' is nothing but saying that `connected events are connected'? If the
non-invariance of the time ordering between the connected events forces us to abandon
such typical conditions on causal relations as the temporal priority of the cause, or
less typical but still reasonable conditions such as simultaneity between cause and effect,
the features of this link are themselves so vague that we should not be worried by the
vagueness of the non-causal terms - namely `dependence, `influence' and the like -
that we might use to denote it: using causal concepts in this case appears then to be a mere
labeling devoid of any real physical and philosophical significance.

\section{Causation and State Reduction}

The options considered so far appear to share an additional drawback, namely that they do not
take into due account the role of the state reduction process. For in EPR-Bell frameworks,
the mark of a relation that we might consider causal between the two outcomes is the
emergence of actual properties of the system on one wing, as a consequence of obtaining
a certain outcome after measuring an observable of the system on the other wing. But in
ordinary quantum mechanics the process itself through which such properties emerge is
exactly the state reduction, so that it is sensible to investigate this notion of
causation at-a-distance yet to be characterized on the background of the reduction process.
Under the assumption that the state reduction is a real physical process (that, as it stands,
lacks Lorentz covariance), there are different options on where the state reduction might take
place and, in view of the above mentioned causation-reduction link, we should take into account
how a notion of causation - even very general - fares with respect with the different accounts
on where the state reduction occurs.

In an early investigation on the non-covariance of the state reduction process, Bloch
argued that the hypersurface on which the state reduction may be taken to occur can be
chosen arbitrarily, since that choice will not affect the probability distribution of
all (local) observables. This prescription is clearly non-covariant, but in a relativistic
quantum theory of measurement ``it appears that either causality or Lorentz covariance of
wave functions must be sacrificed [...] Covariance seems the smaller sacrifice, since
it is apparently not required for the calculation of invariant probabilities.'' ([11], p. 1384).
This argument might provide a motivation for one of the above mentioned options, according
to which it is the very time ordering associated to any Lorentz frame that defines which
is the cause and which the effect. If for instance one performs a measurement in an EPR-Bell
correlation experiment, it can be assumed in the Bloch spirit that the state reduction occurs
along a space-like hyperplane containing the measurement event in the frame of the observer
who performed the measurement. In a later paper Hellwig and Kraus, although still emphasizing
that what matters are just probability distributions since these are insensitive to the Lorentz
frame adopted to order the events, have proposed a prescription according to which the
reduction occurs along the backward light cone of the measurement event ([23]).

In a series of papers Aharonov and Albert have shown that, although Lorentz-covariant,
the Hellwig-Kraus prescription turns out to be inadequate when non-local observables are
taken into account, namely observables of just such composite systems as those considered
in EPR-Bell correlation experiments ([1]-[3]). But also without addressing the Aharonov and
Albert criticisms (the debate is still alive: see for instance [27], [17] and [20]),
the very fact which the Bloch non-covariant prescription and the Hellwig-Kraus covariant one
rely on is unsatisfactory by my specific point of view. Namely, the fact that the expectation
values of the considered observables - be they local or non-local - are invariant across
different Lorentz frames tells us nothing that might be relevant to explaining the
superluminal dependence between single events and perhaps to interpreting it in causal terms.
Moreover, as far as just expectations values are taken into account, quantum mechanics does
satisfy statistical locality in the sense that in a typical EPR-Bell correlation experiment,
for instance, the expectation value of a spin observable pertaining one subsystem is completely
unaffected by any kind of operation performed on the distant subsystem ([16], [21]).
Therefore, should we confine our attention to the level of expectation values, the very
non-locality problem (and the correlated one of attempting a causal interpretation of it)
would not even arise. But there is a further consequence of the Aharonov-Albert analysis that
turns out to be relevant by our viewpoint, namely the revision of the usual meaning ascribed
to the wave function in a relativistic context. According to their proposal, when a local
measurement is performed at a spacetime point S, the state reduction should be taken to
occur along every space-like hyperplane intersecting S. In addition to the Lorentz-covariance
that this proposal allows one to achieve, it implies that the state of the system in a
relativistic quantum-mechanical context must be represented as a functional defined on
the set of space-like hyperplanes, so that in turn the ordinary wave function takes on
different values at a given spacetime point according to which space-like hyperplane is
considered ([3], pp. 231-2). By the point of view of causal relations between events, however,
this implies that certain events - that might play the role of `causes' and that are given
by wave functions taking on definite values at spacetime points - are actual in certain
hyperplanes and not in others. This leads us back to the starting point: also in a
relativistic account of the state reduction process such as the Aharonov-Albert one,
there seem to be no room for a characterization of causation that goes beyond a merely
verbal elaboration of the circumstance that certain events manifest a mutual connection
different in important respects from all other physical forces known in nature.
\footnote{The fact that the ordinary wave function takes on different values at a
given spacetime point according to which space-like hyperplane is considered,
following from the generalization of the state as represented by a functional
on the set of space-like hyperplanes, has analogies with the Fleming hyperplane
dependence approach to quantum states ([18], [19]). It seems to me that the
status of causation in the Fleming approach would be similar to that in the
Aharonov-Albert approach, but this point deserves further investigations.}
More generally, in the shift to relativistic quantum mechanics, there is a circumstance
that Aharonov and Albert emphasize and that seems to be forced upon us by the attempt of
finding a Lorentz-covariant formulation of the measurement process: the theory preserves the
capacity of prescribing the correct probabilities for measurement outcomes, but not the
capacity of attributing definite states to the physical systems whose outcome probabilities
are evaluated. If this is the case, the prospects of a causal view of the relation holding
between the correlated outcome events in EPR-Bell correlation experiments appear rather
dim also in a relativistic quantum-mechanical context: it is problematic to think of EPR-Bell
events as causally connected when these events should be represented as instances of properties
satisfied by the suitable physical systems, but in fact no definite ordinary state can be
attributed to the latter.

\section{Spacetime Foliation, Bohmian Mechanics and Causation}

In ordinary quantum mechanics (i.e. quantum mechanics with state reduction), there would be
in principle a further option that we did not consider so far: a preferred foliation of
spacetime might be explicitly assumed, with respect to which it would be perfectly
determinate which events are causes and which effects. This would amount, however, to a
violation de facto of the above mentioned relativistic constraint, since in this case
the space-like separation between the causally connected events would be only a
phenomenological relation. Moreover, the only reason for such a strong assumption would
be just to make room for space-like causation. This move would also have the somewhat
ironic consequence that, in order to explain a deeply non-classical feature like a
fundamental physical relation between space-like separated events, deeply pre-relativistic
features like the absoluteness of the time ordering between the events themselves are
reintroduced.

The situation is different for Bohmian mechanics, whose overall structure may provide
independent (and deeper) reasons for justifying the assumption of a preferred foliation of
spacetime. As it stands, Bohmian mechanics is not a Lorentz-invariant theory. In the general
case of a $N$-particles system, the guidance equation - the only dynamical law added by the
theory to the Schr\"odinger equation - concerns the positions of the $N$ particles at a
common and absolute time: this presupposes the assumption of a foliation of spacetime
into space-like hyperplanes that, however, turns out to be impossible to determine.
In this context, a causal interpretation of non-locality appears rather natural, since
a causal relation between the EPR-Bell events might be then assumed to hold just with respect
to the foliation. If one is willing to accept that the democracy reigning among Lorentz frames,
prescribed by special relativity, is to be meant just as a phenomenological circumstance
- as David Albert put it, ``taking Bohm's theory seriously will entail being instrumentalist
about special relativity'' ([4], p. 161, emphasis in the original) - then a notion of
space-like causation linking EPR-Bell events may very well be accomodated into Bohmian
mechanics. Admittedly, it would be quite an unconventional sort of causation, since it
would share with the foliation (and with all quantities defined with respect to it) an
epistemic inaccessibility: in this picture, we know that causation is there, although we
are bound to remain ignorant about its mechanisms and about which events are `causes' and
which 'effects'. It is also true, however, that a supporter of Bohmian mechanics need not
being particularly worried by this circumstance, rather disturbing for others. In fact he
should find it relatively easy to accomodate it within the framework of Bohmian mechanics,
since fundamental beables of the theory - in Bell's terminology - like the particles'
positions and trajectories are themselves out of reach. As aptly pointed out by Maudlin,
``if the existence of empirically inaccessible physical facts is fatal, then Bohmian mechanics
is a non-starter even before Relativity comes into play'' ([25], p. 296).

It has been argued that in Bohmian mechanics the essential symmetry of the possibly causal
relations between two typical EPR-Bell subsystems - when for instance the position of one
particle causes the velocity of the other, which is space-like separated from the first -
makes it difficult to speak of a serious `causal' influence between the systems ([13],
p. 325). This argument presupposes that in order for a notion of causal influence to be
meaningful, a direction is to be selected along which the influence is supposed to act.
Once we assume a preferred foliation of spacetime in the spirit of Bohmian mechanics, however,
we need not hold on to a notion of causation in which an event $A$, in order to be causally
related to a different event $B$, must temporally precede $B$ with respect to the preferred
time ordering. We can envisage a sort of causal implication between events similar to that
discussed by Maudlin for ordinary quantum mechanics, the only relevant difference being that
in Bohmian mechanics we can correctly interpret it as a simultaneous and mutual causal
influence, since we have assumed a privileged time ordering.

The logical possibility of admitting causal relations between space-like separated
events, allowed by the foliation implicit in the equations of the theory, need not assume
that `realistic' theories such as Bohmian mechanics must be non Lorentz-invariant. The
argument put forward by Hardy, aiming to show that this is the case ([22]), was soon shown
not to be as compelling as it was thought to be (for a concise review of this discussion,
see [14]). Although it is argued that Lorentz invariance must be violated at the microlevel,
albeit recovered at the statistical level (see e.g. [29]), there is no existing proof that
this is a logical necessity; there are on the contrary some attempts to construct fully
Lorentz invariant models of Bohmian mechanics by the introduction of additional dynamical
structure, that would allow a Lorentz invariant notion of `evolving configuration' along
which superluminal dependences between EPR-Bell events would be transmitted ([25], [10]).
It might be of some interest to investigate the role that a general notion of causation might
play in such models.

\section{Conclusions}

In the previous sections, we have seen that the aim of characterizing an unambiguous notion
of causation linking single space-like separated events in EPR-Bell frameworks is not an easy
one to achieve. A possible reaction to this state of affairs might be to argue that just this
kind of conceptual difficulties supports the thesis according to which a causal view of
non-locality can be sensibly investigated only in a stochastic framework, although the
application of probabilistic theories of causation to the non-locality issue is itself a
controversial matter.

A general conclusion that may be drawn from the discussion above is that, as far as ordinary
quantum mechanics is concerned, we are facing a dilemma: either the notion of causation is
interpreted in such general terms so as to lose sight of the original underlying intuition
- so that we seem to do nothing but giving a different name to the puzzle under scrutiny - or
we are led to ascribe to the special-relativistic spacetime structure a purely phenomenological
status in order to make room for a preferred spacetime foliation, with respect to which causal
relations can be univocally defined. The latter horn, moreover, is in deep tension with the
attempt of constructing a Lorentz invariant account of state reduction, which is supposed
to be exactly the process through which some of the causally relevant properties of physical
systems are actualized. If we move to formulations of quantum theory without state reduction,
standard Bohmian mechanics has at least more serious motivations by a foundational viewpoint
for accepting a violation of Lorentz invariance, and in this case it is no surprise that in a
standard Bohmian framework we can in principle make sense of the notion of a causation at a
distance.

\section*{Acknowledgments.} I wish to thank the audience of the NATO Advanced Research
Workshop in Cracow, and Joseph Berkovitz in particular, for stimulating remarks.
I am also grateful to Mauro Dorato for his comments on an earlier draft of the present paper.

\section*{References}

\begin{enumerate}
\item Aharonov, Y., Albert, D. (1980), ``States and observables in relativistic quantum field
theories'', \emph{Physical Review} D, 3316-3324.
\item Aharonov, Y., Albert, D. (1981) ``Can we make sense out of the measurement process in
relativistic quantum mechanics?'', \emph{Physical Review} D, 359-370.
\item Aharonov, Y., Albert, D. (1984) ``Is the usual notion of time evolution adequate for
quantum-mechanical systems?'' II. Relativistic considerations'', \emph{Physical Review} D, 228-234.
\item Albert, D. (1992) \emph{Quantum Mechanics and Experience}, Harvard University Press.
\item Berkovitz, J. (1998a) ``Aspects of quantum non-locality. I: Superluminal signalling,
action-at-a-distance, non-separability and holism'', \emph{Studies in History and Philosophy of
Modern Physics} 29, 183-222.
\item Berkovitz, J. (1998b) ``Aspects of quantum non-locality. II: Superluminal causation
and relativity'', \emph{Studies in History and Philosophy of Modern Physics} 29, 509-545.
\item Berkovitz, J. (2000a), ``The nature of causality in quantum phenomena'', \emph{Theoria} 15, 87-122.
\item Berkovitz, J. (2000b), ``The many principles of the common cause'', \emph{Reports on
Philosophy} 20, 51-83.
\item Berndl, K., D\"urr, D., Goldstein, S., Zangh\`{\i}, N. (1995), ``EPR-Bell nonlocality,
Lorentz invariance and Bohmian quantum theory'', \emph{Physical Review} A 53, 2062-2072.
\item Berndl, K., D\"urr, D., Goldstein, S., Zangh\`{\i}, N. (1999), ``Hypersurfaces Bohm-Dirac models'',
\emph{Physical Review} A 60, 2729-2736.
\item Bloch, I. (1967) ``Some relativistic oddities in the quantum theory of observation'',
\emph{Physical Review} 156, 1377-1384.
\item Cushing, J., Fine, A. and Goldstein, S. (eds.) (1996), \emph{Bohmian
Mechanics and Quantum Theory: An Appraisal}, Kluwer.
\item Dickson, M. (1996) ``Is the Bohm theory local?'', in [12], 321-330.
\item Dickson, M. (1998), \emph{Quantum Chance and Non-Locality}, Cambridge University Press,
Cambridge.
\item Ducasse, C.J. (1926) ``On the
nature and observability of the causal relation'', \emph{Journal of
Philosophy} 23, 57-68 (reprinted in Sosa, E., Tooley, M. (eds.),
\emph{Causation}, Oxford University Press, 1993, 125-136, page references
to the reprinted edition).
\item Eberhard, P. (1978) ``Bell's theorem
and the different concepts of locality'', \emph{Nuovo Cimento} 46B,
392-419.
\item Finkelstein, J. (2000), ``Property attribution and the
projection postulate in relativistic quantum theory'', preprint
quant-ph/0007105.
\item Fleming, G.N. (1989) ``Lorentz invariant state
reduction and localization'', in Arthur Fine and Jarrett Leplin
(eds.) \emph{PSA1988} Vol. 2, Philosophy of Science Association, 112-126.
\item Fleming, G.N. (1996) ``Just how radical is hyperplane
dependence?'', in R.K. Clifton (ed.), \emph{Perspectives on Quantum
Reality}, Dordrecht: Kluwer, 11-28.
\item Ghirardi, G.C. (2000) ``Local
measurements of nonlocal observables and the relativistic
reduction process'', preprint quant-ph/0003149.
\item Ghirardi, G.C.,
Rimini, A. and Weber, T. (1980) ``A general argument against
superluminal transmission through the quantum mechanical
measurement process'', \emph{Lettere al Nuovo Cimento} 27, 293-298.
\item Hardy, L. (1992) ``Quantum mechanics, local realistic theories and
Lorentz-invariant realistic theories'', \emph{Physical Review Letters} 68,
2981-2984.
\item Hellwig, K.E., Kraus, K. (1970) ``Formal description
of measurements in local quantum field theory'', \emph{Physical Review} D1,
566-571.
\item Maudlin, T. (1994) Quantum Non-Locality and
Relativity, Blackwell, Oxford.
\item Maudlin, T. (1996) ``Space-time
and the quantum world'', in [12], 285-307.
\item Maudlin, T. (2000)
Review of [14], \emph{British Journal for the Philosophy of Science} 51,
875-882.
\item Mould, R.A. (1999), ``A defense of Hellwig-Kraus
reductions'', preprint quant-ph/9912044.
\item Placek, T. (2000), \emph{Is
Nature Deterministic?}, Jagellonian University Press, Cracow.
\item Valentini, A. (forthcoming), \emph{Pilot-Wave Theory of Physics and
Cosmology}, Cambridge University Press, Cambridge.
\end{enumerate}

\end{document}